\documentclass[a4paper,10pt,twoside]{just}

\usepackage{multicol}
\usepackage{graphicx}
\usepackage{booktabs}
\usepackage{amssymb,bm,mathrsfs,bbm,amscd}
\usepackage[tbtags]{amsmath}
\usepackage{lastpage}
\usepackage{CJK}
\begin{document}
\begin{CJK*}{GBK}{song}

\fancyhead[c]{\small  10th International Workshop on $e^+e^-$ collisions from $\phi$ to $\psi$ (PhiPsi15)}
 \fancyfoot[C]{\small PhiPsi15-\thepage}

\title{$\gamma\gamma$ physics and transition form factor measurements
  at KLOE\textbackslash KLOE-2}

\author{%
     Paolo Gauzzi$^{1)}$\email{paolo.gauzzi@roma1.infn.it}
(for the KLOE-2 Collaboration)
}
\maketitle

\address{%
$^1$ Dipartimento di Fisica, Universit\`a La Sapienza e INFN, Sezione di
Roma, Rome, I-00185, Italy\\

}

\begin{abstract}
The KLOE results on the measurement of the transition form factors of the
$\eta$ and $\pi^0$ mesons in $\phi$ Dalitz decays are presented, and the
determination of the $\Gamma(\eta\to\gamma\gamma)$ in $\gamma\gamma$
collisions is also reported.\\
The prospects for $\gamma\gamma$ physics of the new data-taking, started in
November 2014 with the upgraded detector, are reviewed.
\end{abstract}



\begin{multicols}{2}

\section{Introduction}

The KLOE Collaboration took data from 2001 to 2006 at the Frascati
$\phi$-factory DA$\Phi$NE, collecting about 2.5 fb$^{-1}$ at the
peak of the $\phi(1020)$, and 250 pb$^{-1}$ off-peak, mainly at $\sqrt{s}=
1$ GeV. 
In 2008 a new interaction scheme for DA$\Phi$NE has been adopted,
aiming to an increase in luminosity.
Following this successful test, a new data-taking campaign of the KLOE
experiment (KLOE-2 in the following) with an upgraded detector has been
proposed\cite{AmelinoCamelia:2010me}.
The DA$\Phi$NE commissioning for the KLOE-2 data-taking started in 2010.
In December 2012 the machine has been shut down to install the new
beam-pipe with new detectors in KLOE.
In July 2013, after the completion of the installation, the machine
commissioning has been resumed.
The KLOE-2 data taking started in November 2014, with the goal to collect
at least 5 fb$^{-1}$ of integrated luminosity in 2 - 3 years.
Until June 2015 DA$\Phi$NE provided 1 fb$^{-1}$ of luminosity, that has
been collected by KLOE-2 with an efficiency of about 80\%, as shown in
Fig.~\ref{fig:lumi}.
During this period the DA$\Phi$NE peak luminosity was about 2$\times
10^{32}$ cm$^{-2}$ s$^{-1}$, and the integrated luminosity collected in a
day was about 10 pb$^{-1}$.  
\begin{center}
\includegraphics[width=5cm]{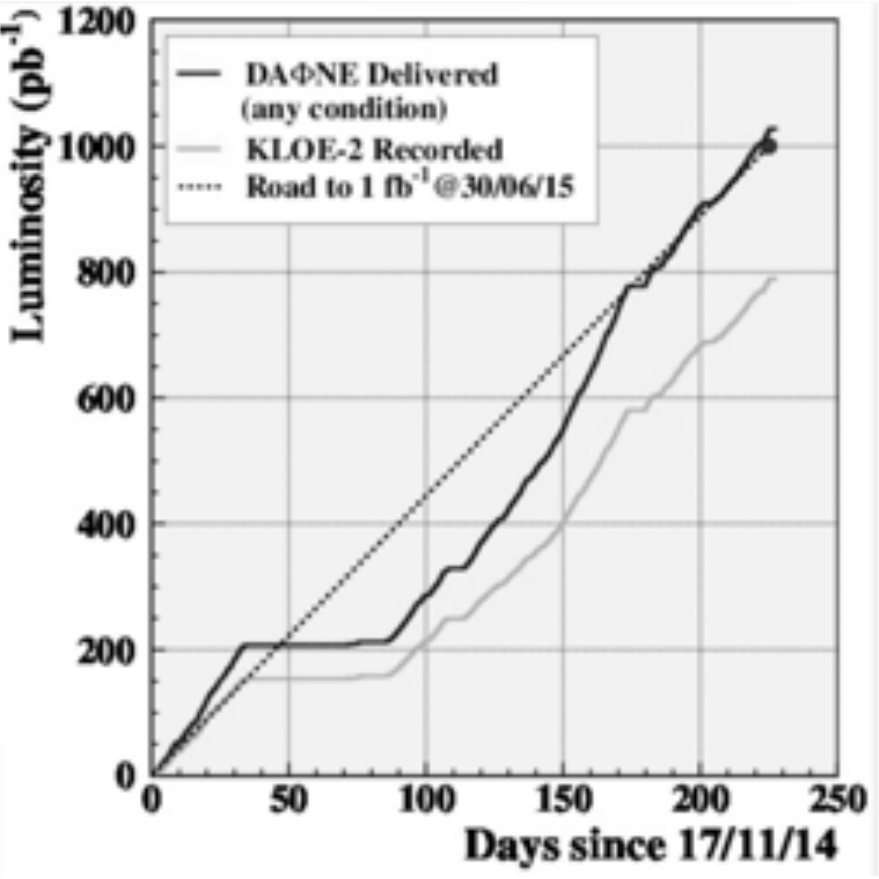}
\figcaption{\label{fig:lumi} Integrated luminosity collected by KLOE-2
  since the start of the new data-taking. }
\end{center}

One of the main items of the KLOE-2 physics
program\cite{AmelinoCamelia:2010me} is the measurement of the Transition
Form Factors (TFFs) of the pseudoscalar mesons both in the space-like and in
time-like region of momentum transfer.
The TFFs describe the coupling of mesons to photons and provide information
about the nature of the mesons and their structure. 
Recently the interest in the TFFs has been renewed since they are an
essential ingredient in the calculation of the hadronic Light-by-Light
(LbL) scattering contribution to the anomalous magnetic moment of the
muon\cite{Jegerlehner:2009ry}.
The leading contribution to the LbL scattering is the single
pseudoscalar exchange (Fig.~\ref{fig:lbl}), where the TFFs enter at the
vertices connecting the pseudoscalar to photons.
\begin{center}
\includegraphics[width=5cm]{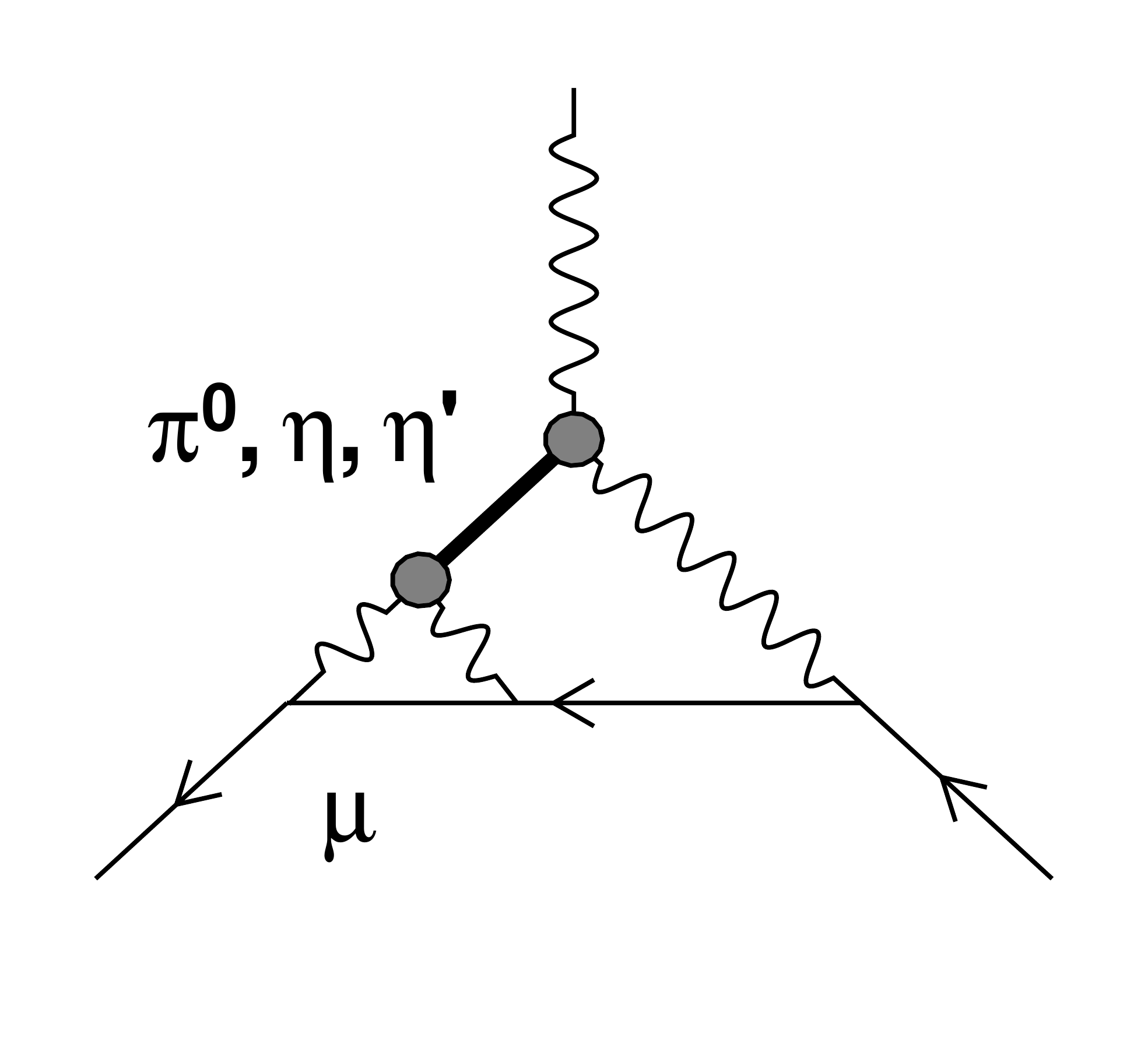}
\figcaption{\label{fig:lbl} Dominant contribution to the hadronic LbL
  scattering for the (g-2)$_{\mu}$ theoretical calculation.}
\end{center}

The calculation of this contribution is model dependent since
the exchanged meson is off-shell, and the TFFs for off-shell meson are not
measurable quantities.
Nevertheless any experimental information, both for space-like and
time-like $q^2$, can help in constraining the models used 
in the calculations. 
The TFFs at time-like $q^2$ can be studied by means of the Dalitz decays,
like $\phi\to\eta e^+e^-$ and $\phi\to\pi^0 e^+e^-$.\\
Another physics item that will be addressed by KLOE-2 is the $\gamma\gamma$
physics, {\it i.e.} processes like
$e^+e^-\to e^+e^-\gamma^\star\gamma^\star\to e^+e^- X$, where $X$ is a final
state with even charge conjugation. 
The expected number of events as a funcion of the $\gamma\gamma$ energy
$W_{\gamma\gamma}$ is:
\begin{equation}
\label{eq:yeld}
\frac{dN}{dW_{\gamma\gamma}} =
L_{int}\frac{dF}{dW_{\gamma\gamma}}\sigma_{\gamma\gamma\to X}
\end{equation}
where $L_{int}$ is the integrated luminosity, $\sigma_{\gamma\gamma\to X}$
the $\gamma\gamma$ cross-section, and $\frac{dF}{dW_{\gamma\gamma}}$ is the
luminosity function which is plotted in Fig.~\ref{fig:efflumi} for three 
different energies.
Since DA$\Phi$NE is operated at $\sqrt{s}\simeq M_{\phi}$, the accessible
final state are either single pseudoscalar, $X=\eta,~\pi^0$, or the double
pion production, $X=\pi\pi$.\\    
The cross-section for single pseudoscalar is:
\begin{equation}
\label{eq:ggxsec}
\sigma_{\gamma\gamma\to
  X}(q^2_1,q^2_2)=\frac{8\pi^2}{M_{X}}\Gamma(X\to\gamma\gamma)\left|F(q^2_1,q^2_2)\right|^2 
\end{equation}
with $(q^2_1+q^2_2)=M^2_X$.

\begin{center}
\includegraphics[width=6cm]{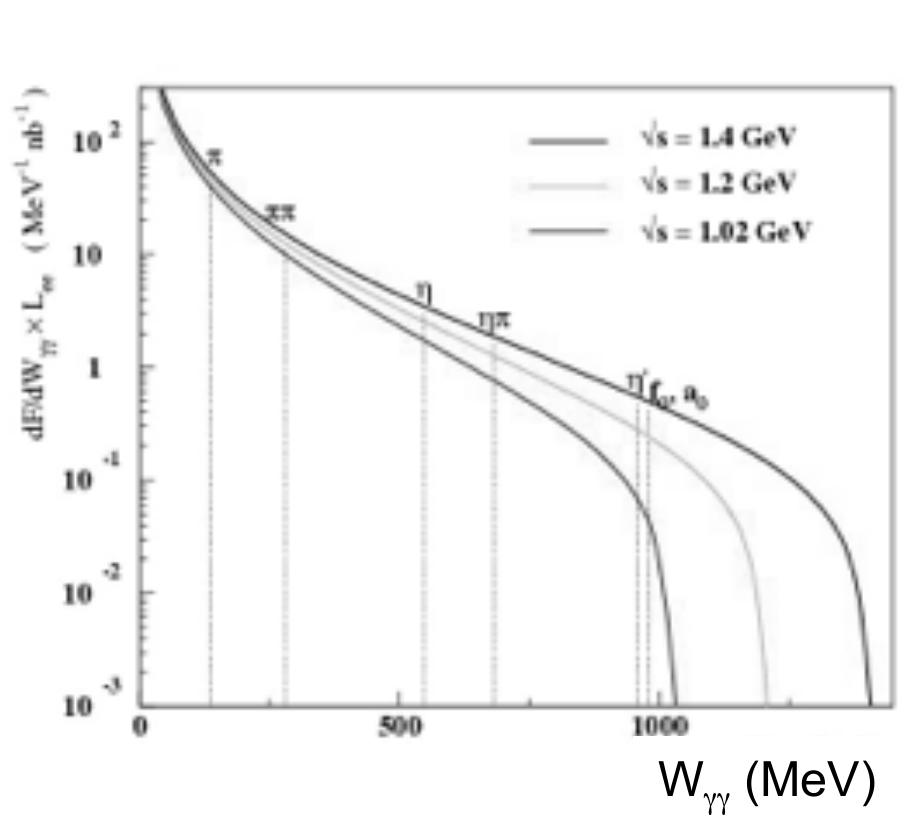}
\figcaption{\label{fig:efflumi} Effective luminosity for
  $\gamma\gamma$ processes for three different energies.}
\end{center}
Then the radiative width $\Gamma(X\to\gamma\gamma)$ of the pseudoscalar
meson, and the TFF $F(q^2_1,q^2_2)$ for space-like $q^2$ can be measured.
Concerning the double pion final state, it is interesting to study the
production of the lowest mass scalar meson $f_0(500)$, but it is also
important for the new dispersive approach proposed for the hadronic LbL
scattering\cite{Colangelo:2015ama}.\\ 

\section{Detector upgrade}

As a first step of the detector upgrade, a tagger system for scattered
electrons and positrons in $\gamma\gamma$ processes has been installed
already in 2010.
It consists of two different devices: the Low Energy Tagger (LET) and the
High Energy Tagger (HET), referring to the energy of the detected electrons
or positrons (see Fig. ~\ref{fig:taggers}).\\ 
\begin{center}
\includegraphics[width=6cm]{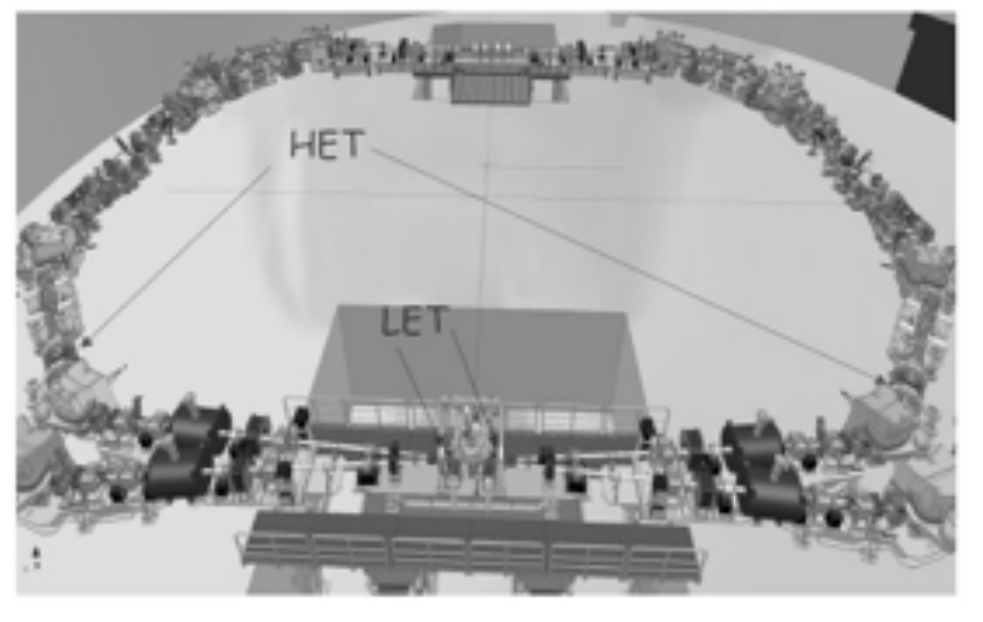}
\figcaption{\label{fig:taggers} Positioning of the taggers along the
  DA$\Phi$NE ring.}
\end{center}

During the 2013 shutdown an Inner Tracker\cite{Balla:2014kqa} made of four
layers of cylindrical triple GEM has been installed between the beam-pipe and the  
Drift Chamber, to improve the resolution for decay vertices close to the
interaction point (IP), and to increase the acceptance for low momentum tracks. 
Furthermore two Crystal Calorimeters (CCALT)\cite{Cordelli:2013mka} have
been added to cover the low polar angle regions to increase the acceptance
for photons and $e^{\pm}$, originating from the IP, down to $10^{\circ}$,
and finally the DA$\Phi$NE focusing quadrupoles, that are placed inside the
KLOE detector, have been instrumented with calorimeters
(QCALT)\cite{Cordelli:2009xb} made of tungsten and scintillator tiles. \\

\subsection{The Low Energy Tagger}

The LET\cite{Babusci:2009sg} has been designed to detect $e^{\pm}$ with energy between 150 and
350 MeV escaping from the beam-pipe, and it is placed at about 1 m from the IP.
Since in this region there is no correlation between the energy and the
scattering angle of the particles, a calorimetric device has been choosen.
Then the LET consists of two calorimeters, each made of 4 $\times$ 5 LYSO
crystals of 1.5 $\times$ 1.5 $\times$ 20 cm$^3$ dimensions.
The crystals are readout by SiPM.
The two calorimeters are placed simmetrically with respect to the IP, as
shown in Fig.~\ref{fig:let}.
\begin{center}
\includegraphics[width=6cm]{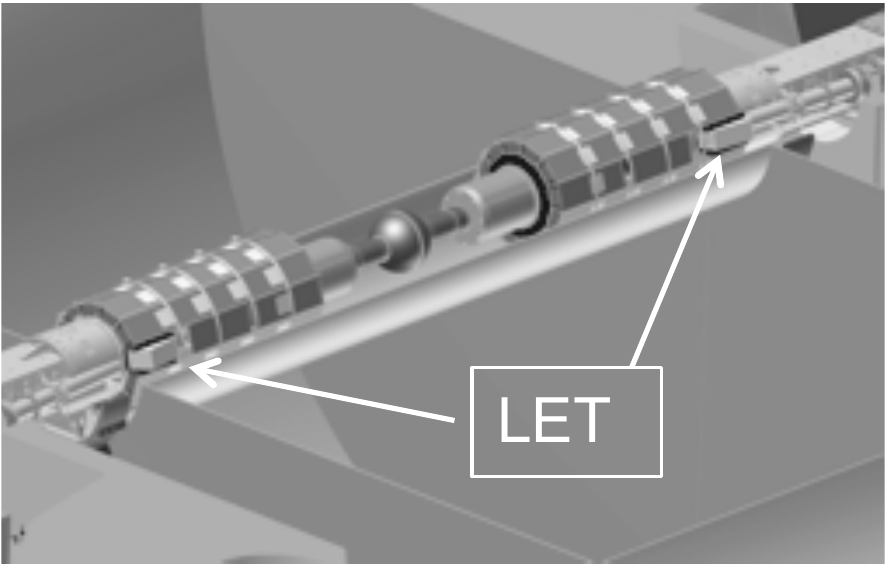}
\figcaption{\label{fig:let} Sketch of the LET calorimeter positioning.}
\end{center}

\subsection{The High Energy Tagger}

The HET\cite{Archilli:2010zza} is designed to detect scattered $e^{\pm}$ of $E > 400$ MeV.
These particles escape the beam-pipe after the first bending dipole of
DA$\Phi$NE, that can thus be used as spectrometer.
The trajectories of the scattered electrons are strongly correlated with
their energy.
Then the HET is made of two scintillator hodoscopes readout by PMT,
symmetrically placed 11 m far from the IP(Fig.~\ref{fig:het}).
\begin{center}
\includegraphics[width=6cm]{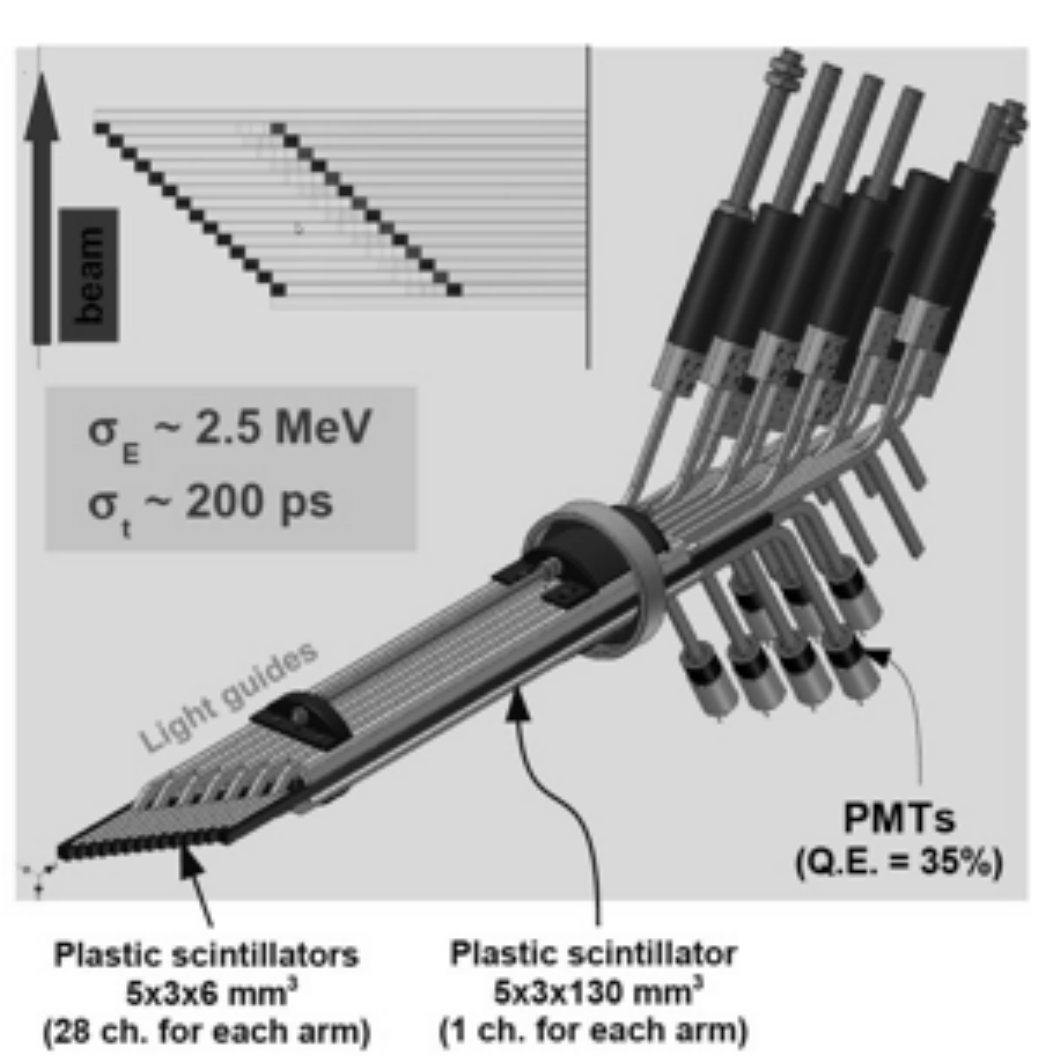}
\figcaption{\label{fig:het} Sketch of the HET hodoscope.}
\end{center}

The HET is acquired asynchronously with respect to the main KLOE detector,
and for each KLOE trigger the HET information concerning three DA$\Phi$NE
beam revolutions is stored.
The synchronization is performed by using a machine signal.
In Fig.~\ref{fig:dthet} the time difference between the two HET stations is
shown: the accelerator time structure of about 2.7 ns period is clearly visible.
The superimposed histogram is the same distribution resulting from a run
with separated beams in the IP, and shows that the level of background is
less than 10\%.\\ 
\begin{center}
\includegraphics[width=6cm]{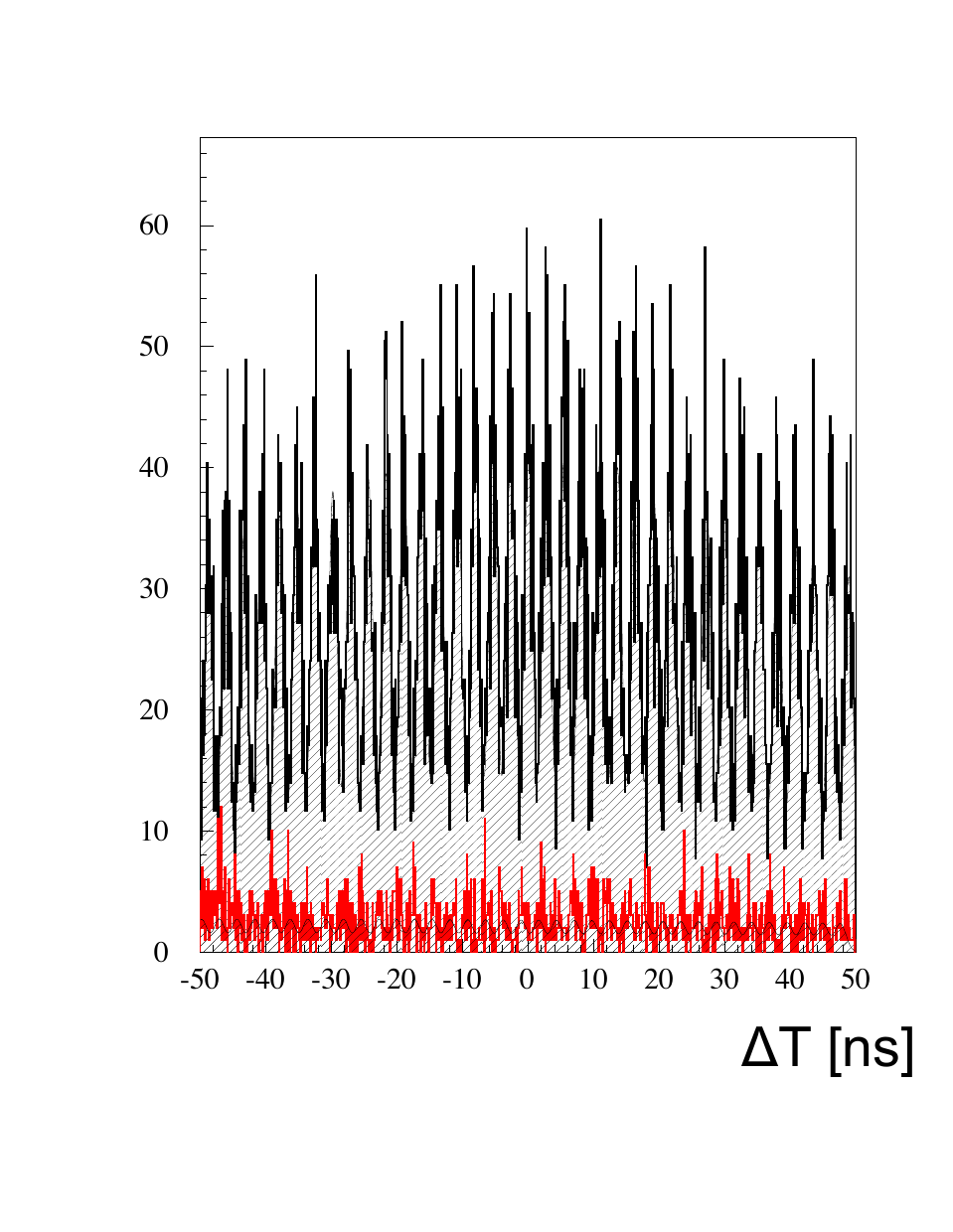}
\figcaption{\label{fig:dthet} Time difference between the two HET
  statstions. Black: colliding beams; red: no collisions.}
\end{center}

\section{$\gamma\gamma$ physics wihout taggers}

With the KLOE data the two-photon width of the $\eta$ meson has been
measured by detecting events $e^+e^-\to e^+e^-\eta$, with $\eta\to\pi^+\pi^-\pi^0$ and
$\pi^0\pi^0\pi^0$. 
The scattered leptons were not detected because the taggers were not
present, then in order to avoid the large background from $\phi$ decays the
data collected off-peak, at $\sqrt{s}=1$ GeV, have been
analyzed, corresponding to an integrated luminosity of 250 pb$^{-1}$. 
In Fig.~\ref{fig:ggeta1}-\ref{fig:ggeta2} the distributions of the
missing mass with respect to $\pi^+\pi^-\pi^0$ and $\pi^0\pi^0\pi^0$,
respectively, are shown.
By fitting these histograms we obtained the cross sections
$\sigma(e^+e^-\to e^+e^-\eta)=(34.5 \pm 2.5 \pm 1.3)~{\rm pb}$ and
$\sigma(e^+e^-\to e^+e^-\eta)=(32.0 \pm 1.5 \pm 0.9)~{\rm pb}$ for the
charged and neutral $\eta$ decay channel, respectively. 
By combining them, $\sigma(e^+e^-\to e^+e^-\eta)=(32.7 \pm 1.3 \pm
0.7)~{\rm pb}$, from which we extract the most precise measurement to date
of the two-photon width: $\Gamma(\eta\to\gamma\gamma)=(520 \pm
20 \pm 13)~ {\rm eV}$\cite{Babusci:2012ik}. \\
\begin{center}
\includegraphics[width=6cm]{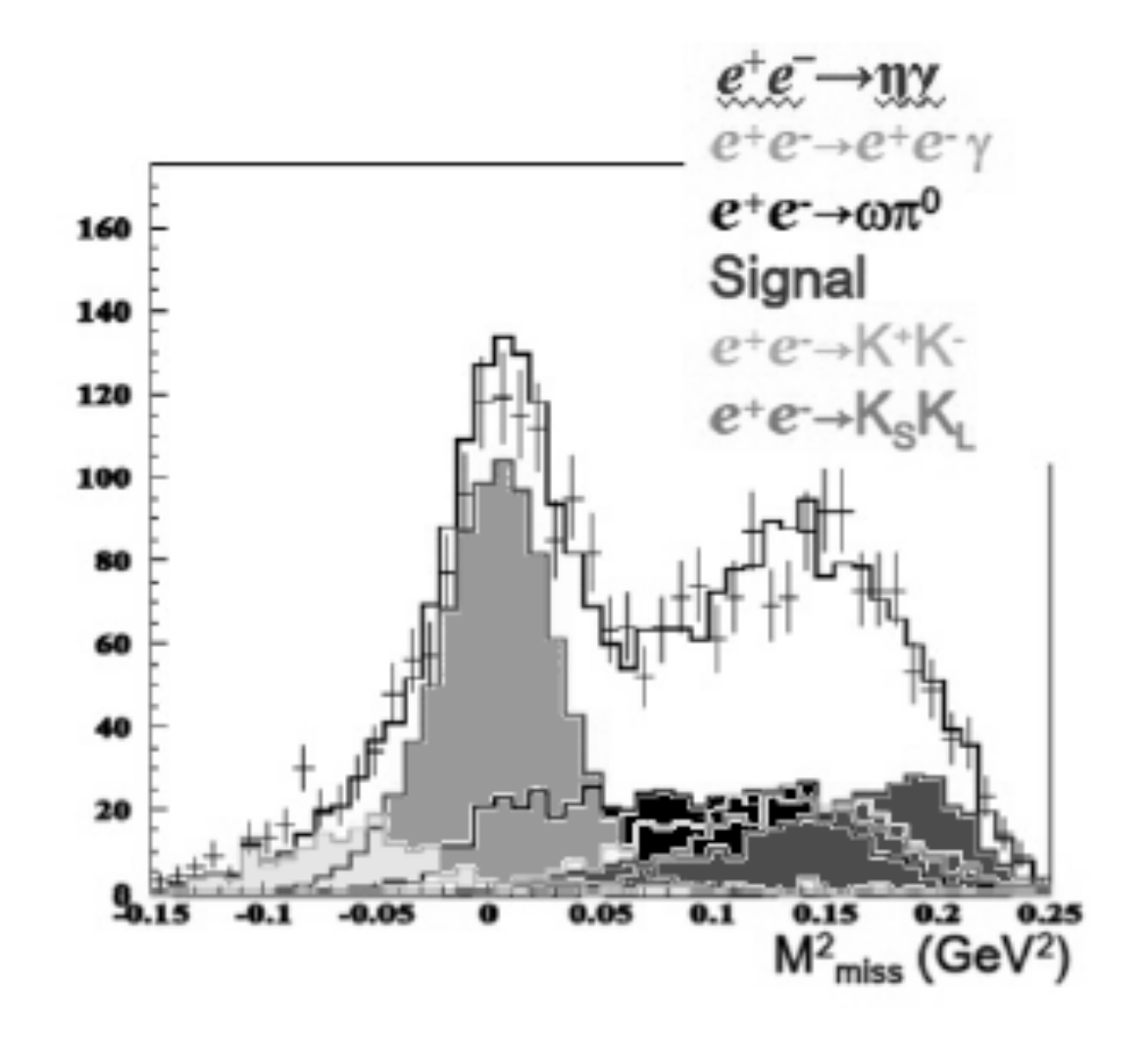}
\figcaption{\label{fig:ggeta1}$\gamma\gamma\to\eta\to\pi^+\pi^-\pi^0$.}
\end{center}

\begin{center}
\includegraphics[width=6cm]{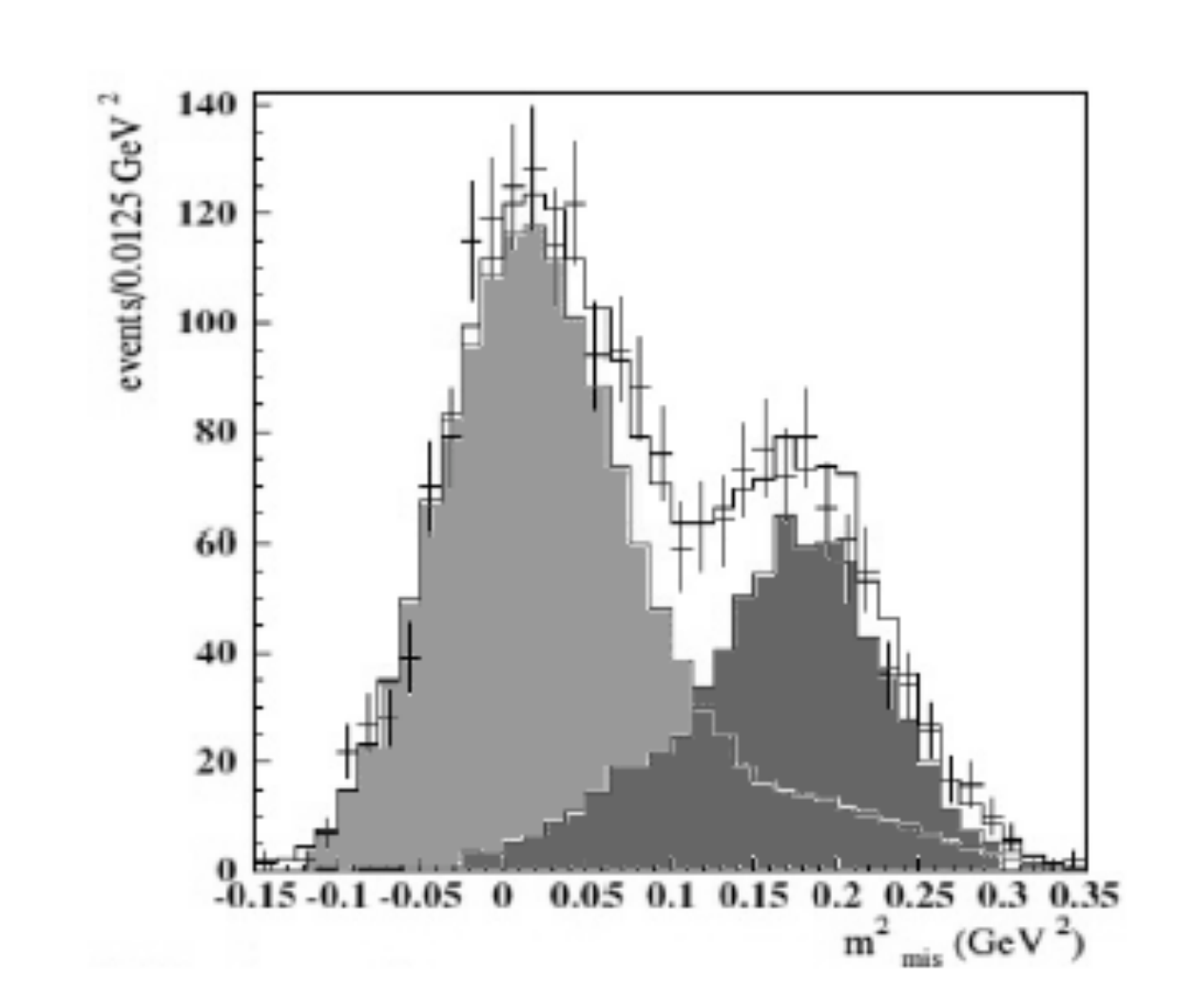}
\figcaption{\label{fig:ggeta2} Dark grey: $\gamma\gamma\to\eta\to\pi^0\pi^0\pi^0$;
  light grey: $e^+e^-\to\eta\gamma$ with  $\eta\to\pi^0\pi^0\pi^0$.}
\end{center}

\section{Prospects for $\gamma\gamma$ physics with taggers}

Since the KLOE-2 data-taking is performed at the $\phi$ peak, the detection
of the scattered electrons and positrons in the tagging stations will be
essential for closing the kinematics of the events and thus to reduce the large
background coming from $\phi$ decays.

\subsection{$\gamma\gamma\to\pi^0$}

The radiative width of the $\pi^0$ has been calculated in Chiral
Perturbation Theory with 1.4\% uncertainty,
$\Gamma(\pi^0\to\gamma\gamma)=(8.09\pm 0.11)$ eV\cite{Kampf:2009tk}, and from the 
experimental point of view, the most precise measurement up to now comes
from the PrimEx Collaboration and is based on the Primakoff effect,
$\Gamma(\pi^0\to\gamma\gamma)=(7.82\pm 0.14\pm 0.17)$ eV\cite{Larin:2010kq}.
However the measurements based on Primakoff effect suffer from some model
dependence due to the conversions in the nucleus field. 
At KLOE-2 the $\pi^0$ width can be measured with a different process by
selecting $e^+ e^-\to e^+ e^-\pi^0$ events with quasi-real photons ($q^2\simeq 0$).
These events are selected by requiring that the scattered $e^{\pm}$ go in
the two HET stations, and the two photons from $\pi^0$ decay are detected
in the calorimeter.
According to the Monte Carlo (MC) simulation the double HET coincidence efficiency is
1.4\%, then for a cross-section $\sigma(e^+ e^-\to e^+ e^-\pi^0)=0.28$ nb,
about 2000 events/fb$^{-1}$ are expected, allowing to reach a 1\% accuracy in
$\Gamma(\pi^0\to\gamma\gamma)$ with 5 fb$^{-1}$ of integrated luminosity.\\  
Moreover the $\pi^0\gamma^*\gamma$ TFF with a quasi-real photon and a
virtual one can also be measured, by selecting events in which one
electron is detected in the HET ($q^2\simeq 0$) and the other one at large
angle in the KLOE main detector.  
In this way a still unexplored $q^2$ region ($|q^2|<0.1$ GeV$^2$, see
Fig.~\ref{fig:ggpi0}), which is important to constrain the TFF
parametrizations, can be investigated. \\ 
\begin{center}
\includegraphics[width=6cm]{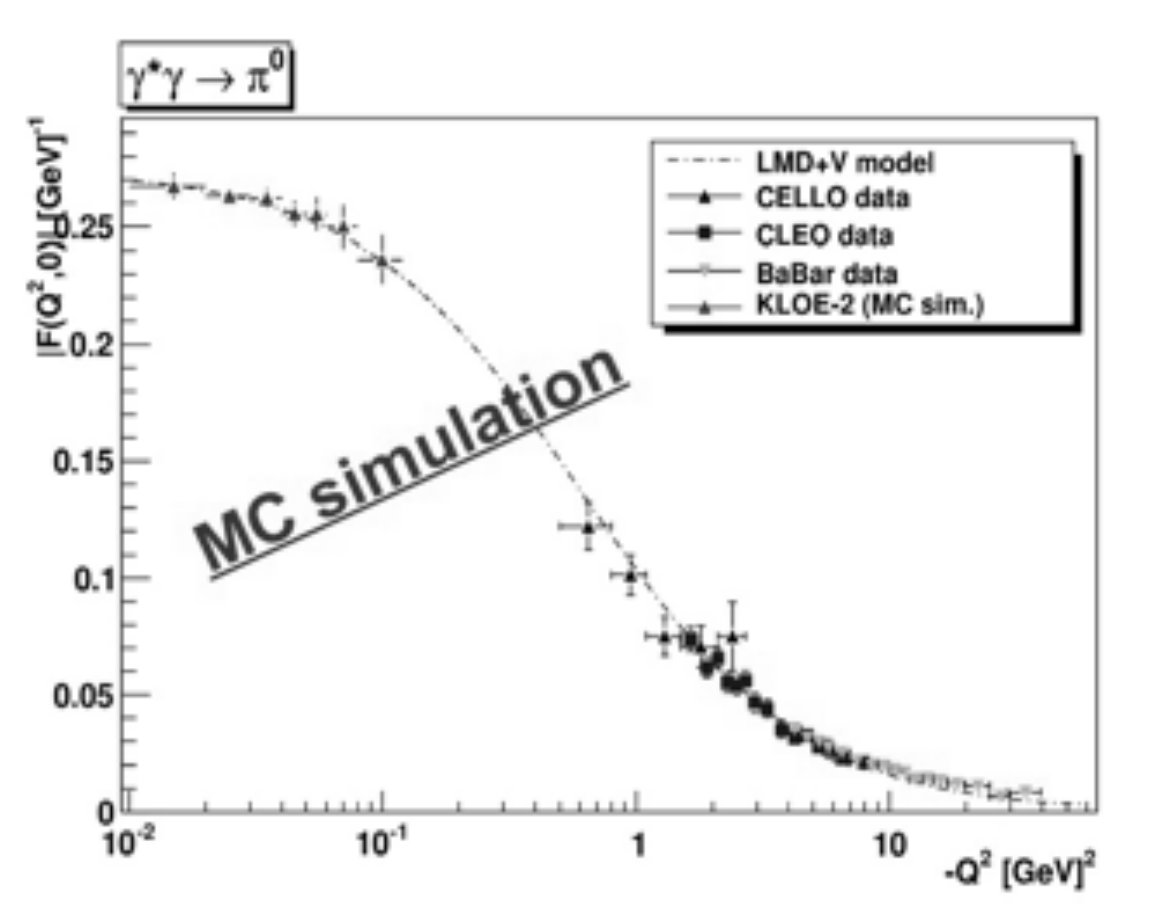}
\figcaption{\label{fig:ggpi0} TFF as a function of $q^2$. The KLOE-2 points
  are from a MC simulation.}
\end{center}

\subsection{$\gamma\gamma\to\pi^0\pi^0$}

In Fig.~\ref{fig:ggpi0pi0} is shown the four photon invariant mass
distribution from a preliminary analysis performed on the old KLOE data
sample: there is a clear excess of events, with respect to all known
background sources, at low mass values. 
\begin{center}
\includegraphics[width=6cm]{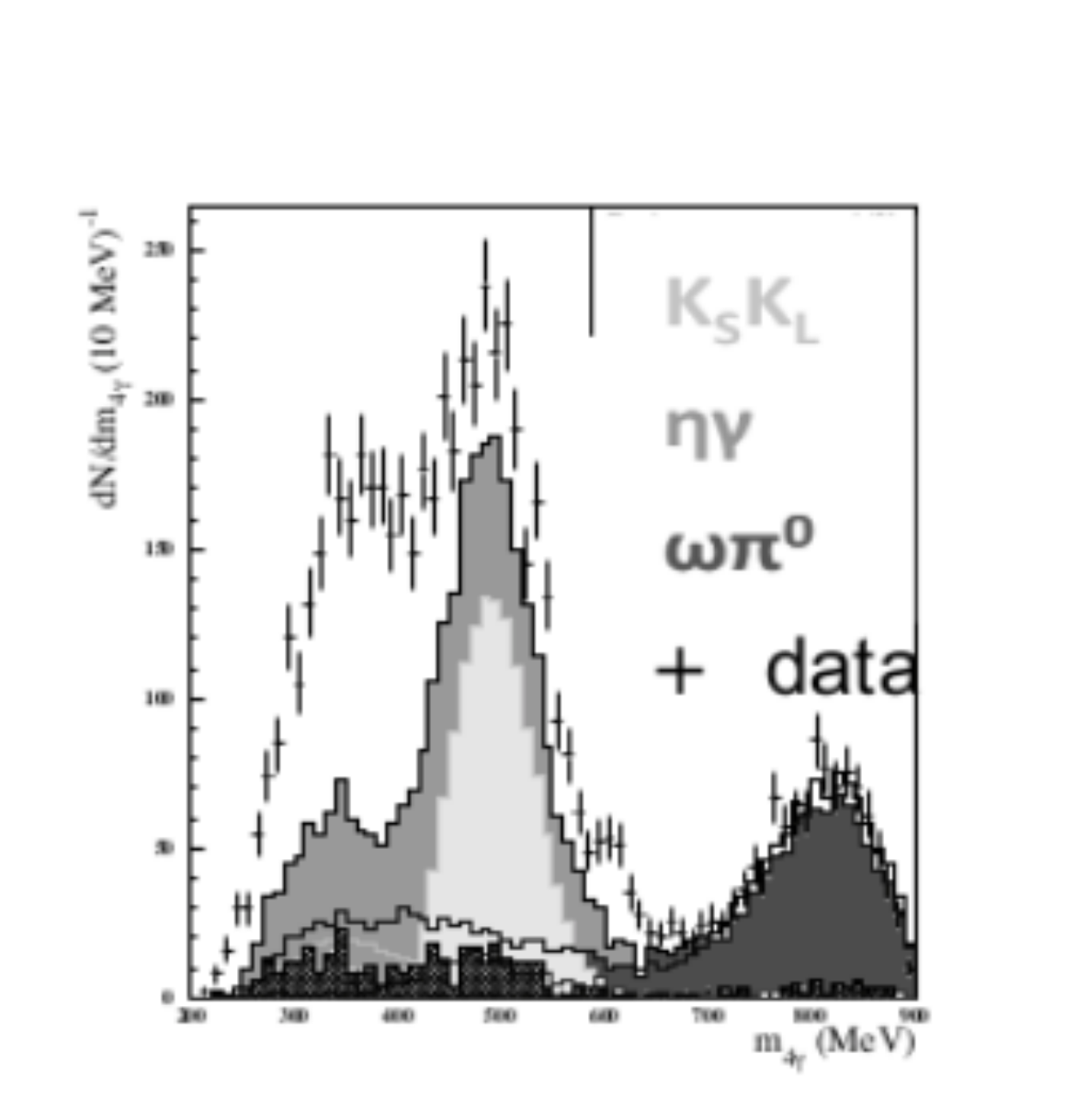}
\figcaption{\label{fig:ggpi0pi0} Off-peak KLOE data sample: four photon
  invariant mass for $e^+e^-\to e^+e^-\pi^0\pi^0$, the top solid histogram
is the sum of all the background processes.}
\end{center}

However, since it is impossible to close the kinematics due to the absence
of the taggers, the residual background is of difficult evaluation.
The measurement of the $e^+ e^-\to e^+e^-\pi^0\pi^0$ cross section will be
possible with the KLOE-2 data, where the relevant energy region can be
covered by selecting events with either LET-LET or HET-LET coincidences.\\

\section{Transition form factor measurements in Dalitz decays}

The TFFs measured in Dalitz decays are functions of the four-momentun squared
$q^2=m^2_{\ell^+\ell^-}$, and according to Vector Meson Dominance (VMD) are
usually parametrized as $F(q^2)=1/(1-\frac{q^2}{\Lambda^2})$, where
$\Lambda$ is a characteristic mass, identified with the nearest vector meson.   
The dilepton invariant mass distributions of $\eta\to e^+
e^-\gamma$ and $\eta\to\mu^+\mu^-\gamma$, measured by NA60
\cite{Uras:2011zz} and by the A2 Collaboration at
MAMI\cite{Berghauser:2011zz}\cite{Aguar-Bartolome:2013vpw} are described  by
$\Lambda^{-2}_{\eta} = 1.92\div 1.95$ GeV$^{-2}$ in agreement with 
the VMD predictions $\Lambda^{-2}_{\eta} = 1.88$ GeV$^{-2}$, while the TFF
of $\omega\to\pi^0\mu^+\mu^-$, also measured by NA60, is not well reproduced by VMD,  
$\Lambda^{-2}_{\omega} = 2.24$ GeV$^{-2}$, while the VMD expectation is
1.68 GeV$^{-2}$. 
To explain this behaviour other models have been
proposed\cite{Terschlusen:2011pm}\cite{Schneider:2012ez}\cite{Ivashyn:2011hb},
that predict deviations from VMD also for $\phi\to\eta(\pi^0)\ell^+\ell^-$.\\ 

\subsection{$\phi\to\eta e^+e^-$}

The TFF slope for $\phi\to\eta e^+e^-$ was measured with low statistics by
the SND Collaboration at Novosibirsk, $\Lambda^{-2}_{\phi} = (3.8 \pm 1.8)$
GeV$^{-2}$\cite{Achasov:2000ne}.
This value is compatible, due to its large uncertainty, with
the VMD expectation $\Lambda^{-2}_{\phi}\simeq m^{-2}_{\phi}\simeq 1$ GeV$^{-2}$.
At KLOE 1.7 fb$^{-1}$ of data have been analyzed looking for $\phi\to\eta e^+ e^-$ with
$\eta\to\pi^0\pi^0\pi^0$.
The $m_{e^+ e^-}$ distribution is shown in Fig.~\ref{fig:trff}.
\begin{center}
\includegraphics[width=6cm]{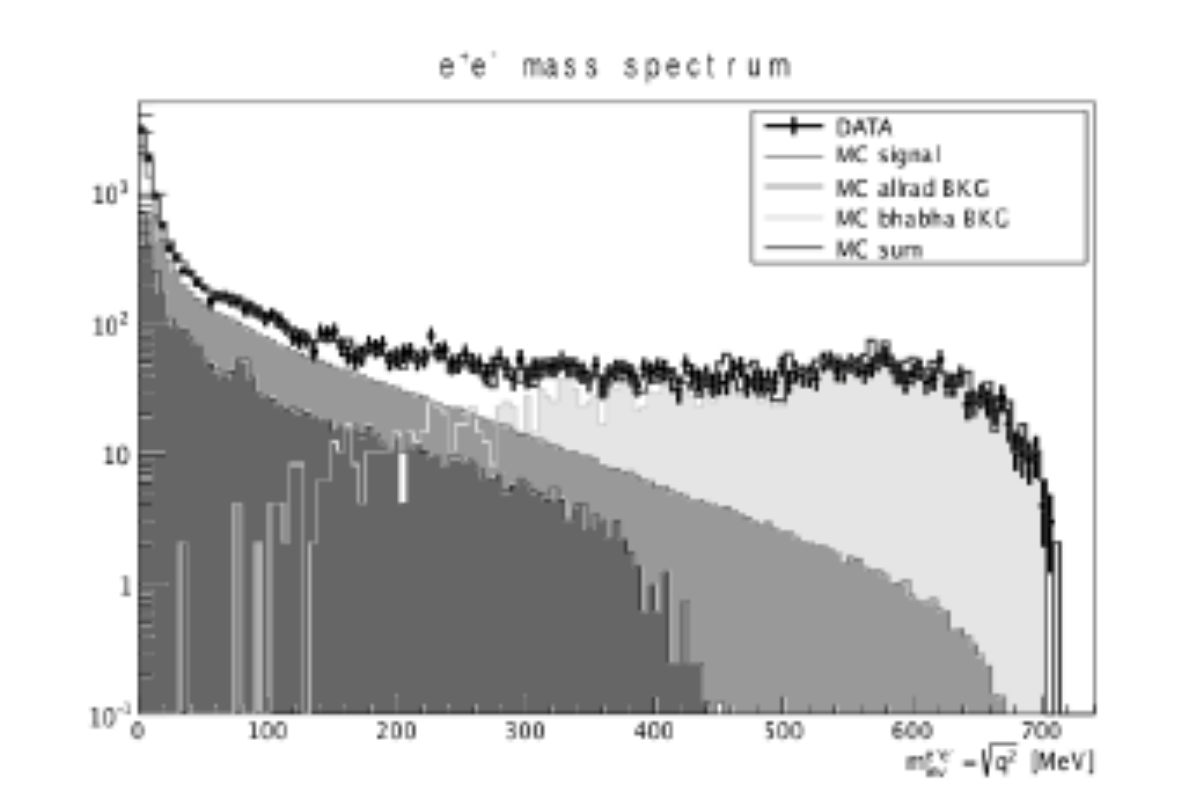}
\figcaption{\label{fig:trff} $e^+e^-$ invariant mass for $\phi\to\eta e^+e^-$.}
\end{center}

From the event counting the branching ratio can be obtained:
$Br(\phi\to\eta e^+e^-)=(1.075 \pm 0.007 \pm 0.038)\times 10^{-4}$\cite{Babusci:2014ldz}. 
The slope $b=\Lambda^{-2}_{\phi}$ is then extracted from a fit of the distribution
of the $e^+e^-$ invariant mass to the parametrization from ref.\cite{Landsberg:1986fd},
by using the one-pole formula for the TFF.
We obtain a value, $b=(1.17 \pm 0.10^{+0.07}_{-0.11})$ GeV$^{-2}$\cite{Babusci:2014ldz}, which is
consistent with the VMD predictions. 
The TFF as a function of the $e^+e^-$ invariant mass is shown in Fig.~\ref{fig:ffvsm}.\\
\begin{center}
\includegraphics[width=6cm]{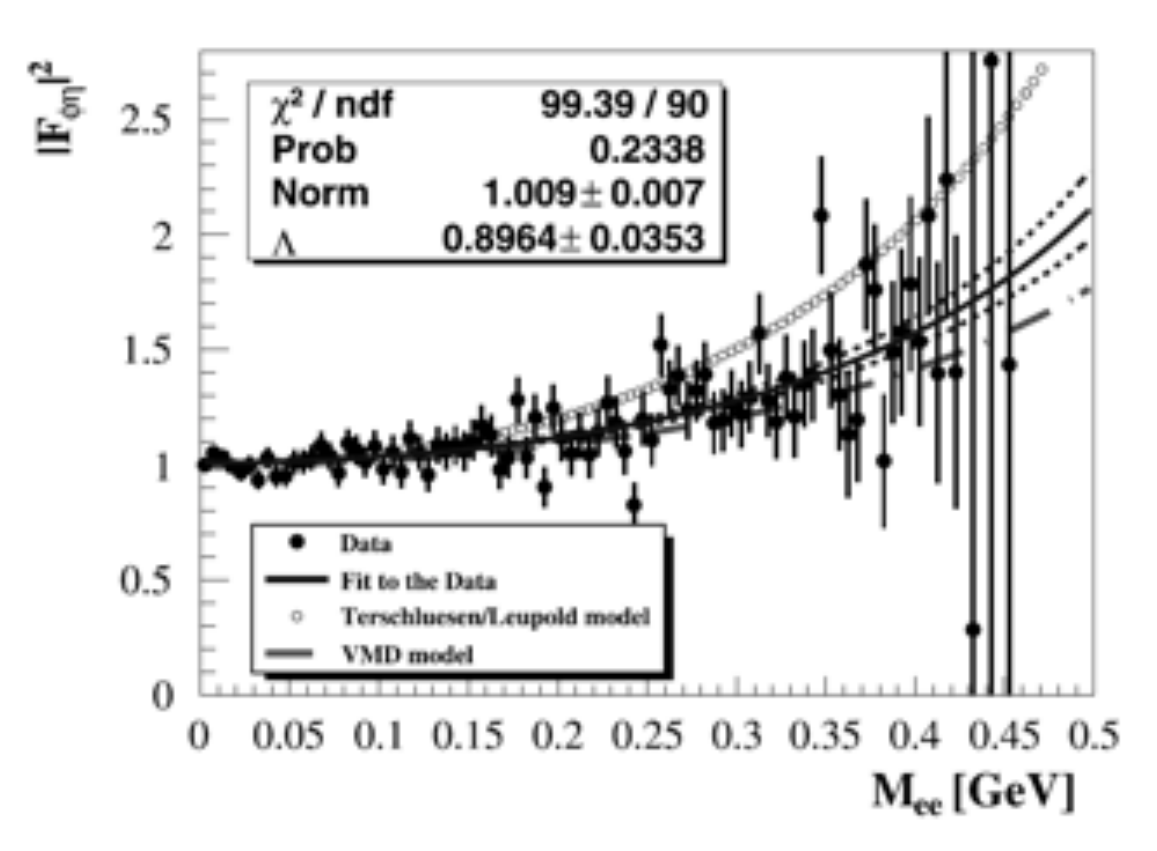}
\figcaption{\label{fig:ffvsm} TFF as a function of the $e^+e^-$ invariant
  mass, compared with different theoretical predictions.}
\end{center}

\subsection{$\phi\to\pi^0 e^+e^-$}
 
The decay $\phi\to\pi^0 e^+e^-$ has been studied by the Novosibirsk
experiments CMD-2 and SND, that reported $Br(\phi\to\pi^0 e^+e^-)=(1.22\pm
0.34\pm 0.21)\times 10^ {-5}$\cite{Akhmetshin:2000wi} and $(1.01\pm 0.28\pm
0.29)\times 10^{-5}$\cite{Achasov:2002hs}, respectively, but no measurement
has been published on the TFFs slope.\\ 
In the sample of 1.7 fb$^{-1}$ of KLOE data, about 9000 events for this
decay have been selected.
In Fig.~\ref{fig:ppee} the data-MC comparison is shown for the $e^+e^-$ 
invariant mass, and in Fig.~\ref{fig:mgg} for the two photon invariant mass.
The residual background, mainly coming from radiative Bhabha scattering, is
subtracted by fitting the distribution of the recoil mass against the
$e^+e^-$ pair.\\
\begin{center}
\includegraphics[width=6cm]{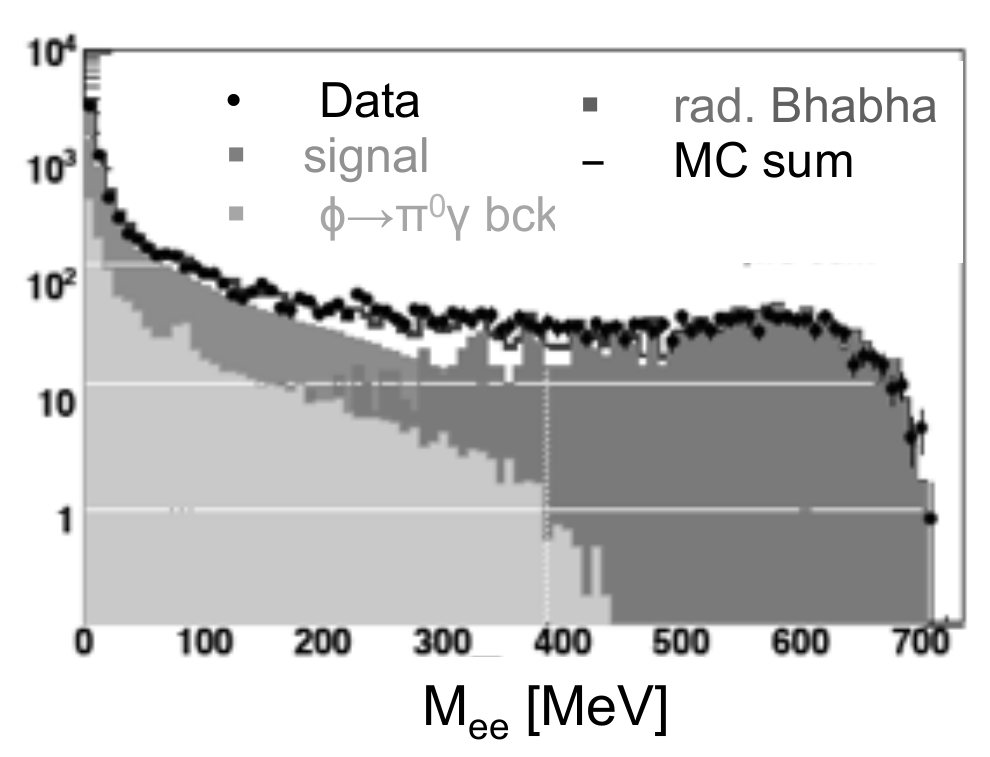}
\figcaption{\label{fig:ppee} $e^+e^-$ invariant mass distribution for $\phi\to\pi^0 e^+e^-$.}
\end{center}

\begin{center}
\includegraphics[width=6cm]{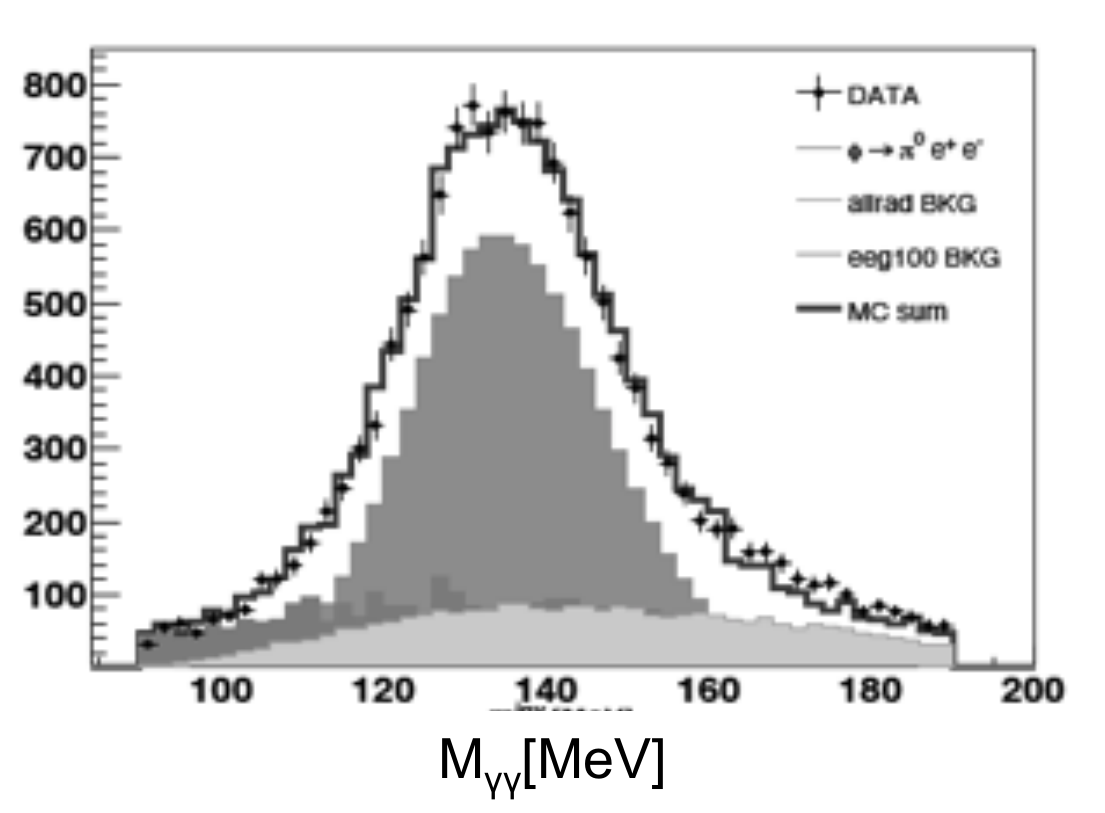}
\figcaption{\label{fig:mgg} Two photon invariant mass distribution for $\phi\to\pi^0 e^+e^-$.}
\end{center}
 
The TFF as a function of the $e^+e^-$ invariant mass is obtained after the
background subtraction and in Fig.~\ref{fig:tffpi0} is compared with the
theoretical expectations.
It shows a good agreement with the model of ref.\cite{Ivashyn:2011hb}.
From the subtracted invariant mass spectrum the branching ratio also can be
derived, $Br(\phi\to\pi^0 e^+e^-)=(1.19\pm 0.05^{+0.05}_{-0.10})\times 10^{-5}$ for
$m_{ee}<700$ MeV.
Higher values of the invariant mass are not accessible due to the selection
of the events; however the branching ratio for all the invariant mass range
can be extrapolated according to the model of ref.\cite{Ivashyn:2011hb},
$Br(\phi\to\pi^0 e^+e^-)=(1.35\pm 0.05^{+0.05}_{-0.10})\times 10^{-5}$. \\

\begin{center}
\hspace{-2cm}\includegraphics[width=9cm]{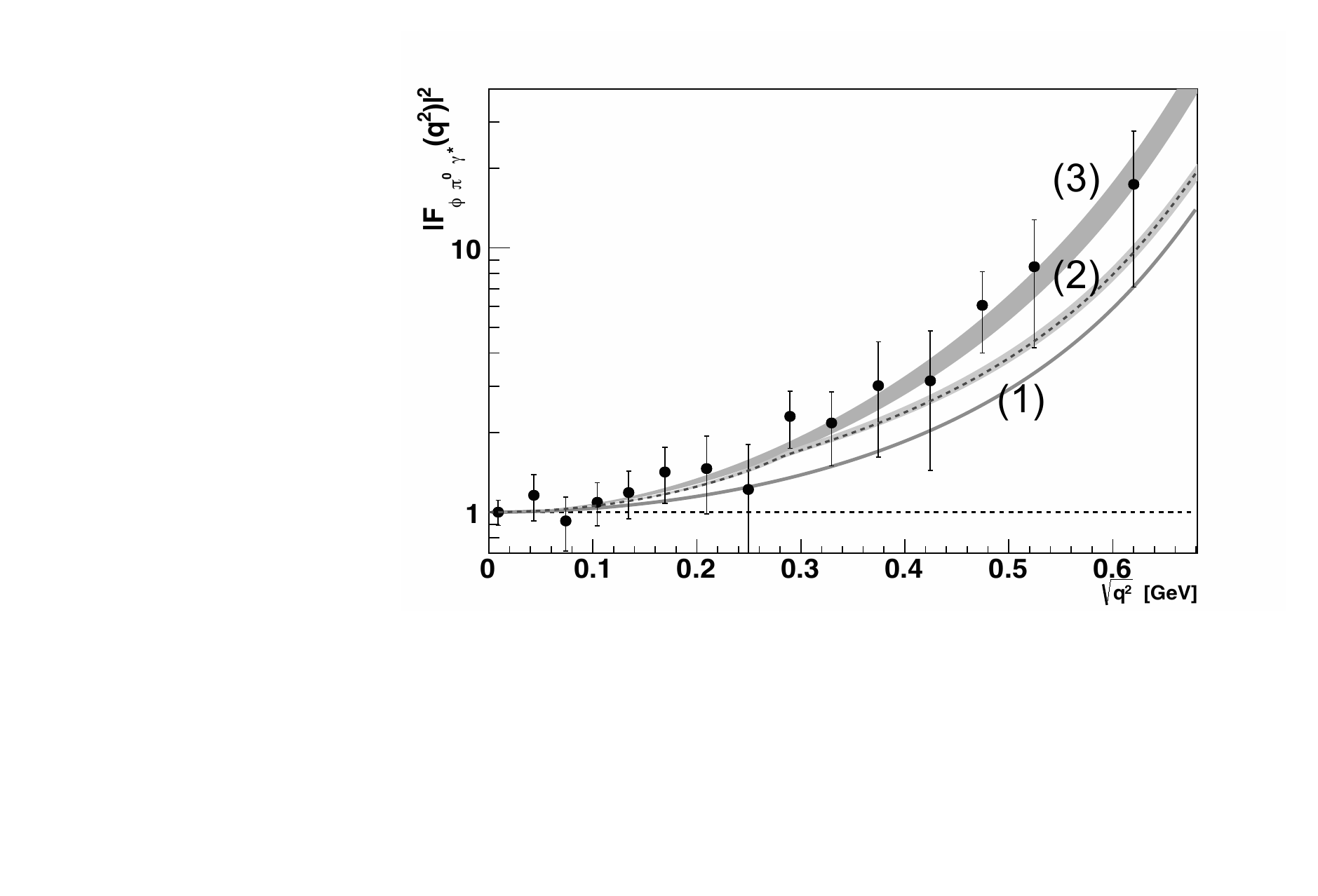}
\vspace{-1cm}\figcaption{\label{fig:tffpi0} $\phi\to\pi^0 e^+e^-$: TFF as a function of
  $e^+e^-$ invariant mass, compared with the following theoretical
  predictions, (1):VDM, (2)-grey band: ref.\cite{Schneider:2012ez},
  (2)-dashed line: ref.\cite{Danilkin:2014cra}, (3): ref.\cite{Ivashyn:2011hb}.}  
\end{center}

\section{Conclusions}
The KLOE Collaboration is continuing to exploit the high
statistics sample of light mesons collected during the first phase of the
experiment, to perform precision measurements in hadron physics.
The most accurate measurement up to now of the $\eta$ width in
$\gamma\gamma$, $\Gamma(\eta\to\gamma\gamma)=(520\pm
20\pm 13)~ {\rm eV}$, has been obtained.\\
From the study of the $\phi$ Dalitz decays, the branching ratios and the
Transition Form Factors of $\phi\to\eta e^+e^-$ and
$\phi\to\pi^0 e^+e^-$ have been measured.\\ 
The KLOE-2 data-taking, with the upgraded detector, is in progress, with
the goal to collect at least 5 fb$^{-1}$ in the next years.
A rich program of measurements has been proposed\cite{AmelinoCamelia:2010me}. 
Among these measurements a special place, due to the new taggers, is
occupied by the $\gamma\gamma$ production of pseudoscalar mesons, that can
help to shed light on some of the still puzzling questions in this field.

\end{multicols}

\vspace{-1mm}
\centerline{\rule{80mm}{0.1pt}}
\vspace{2mm}

\begin{multicols}{2}
\bibliography{pgauzzi_phipsi15}


\end{multicols}

\clearpage

\end{CJK*}
\end{document}